\documentclass[sigconf]{acmart}

\usepackage{balance}

\def\BibTeX{{\rm B\kern-.05em{\sc i\kern-.025em b}\kern-.08emT\kern-.1667em\lower.7ex\hbox{E}\kern-.125emX}}
    
\usepackage{booktabs} 
\usepackage{amsmath,amssymb,amsfonts}
\usepackage{algorithm}
\usepackage{algorithmic}
\usepackage{graphicx}
\usepackage{textcomp}
\usepackage{xcolor}
\usepackage{multirow}
\usepackage{subfigure}
\usepackage[super]{nth}

\graphicspath{ {img/} }

\def\bW{\mathbf{W}}
\def\bA{\mathbf{A}}

\def\bY{\mathbf{Y}}
\def\bK{\mathbf{K}}

\def\bw{\mathbf{w}}
\def\bb{\mathbf{b}}
\def\be{\mathbf{e}}

\def\bd{\mathbf{d}}
\def\bh{\mathbf{h}}

\def\be{\mathbf{e}}
\def\bf{\mathbf{f}}
\def\bu{\mathbf{u}}
\def\bq{\mathbf{q}}

\def\bx{\mathbf{x}}
\def\by{\mathbf{y}}
\def\bh{\mathbf{h}}

\def\bs{\mathbf{s}}
\def\bu{\mathbf{u}}

\def\br{\mathbf{r}}

\DeclareMathOperator*{\argmax}{argmax}

\begin{document}

\copyrightyear{2019} 
\acmYear{2019}  
\setcopyright{acmcopyright}
\acmConference[KDD '19]{The 25th ACM SIGKDD Conference on Knowledge Discovery and
Data Mining}{August    4--8, 2019}{Anchorage, AK, USA}
\acmPrice{15.00}
\acmDOI{10.1145/3292500.3330965}
\acmISBN{978-1-4503-6201-6/19/08}

\title{ 
Beyond Personalization: Social Content Recommendation \\ for Creator Equality and Consumer Satisfaction 
}


\author{Wenyi Xiao\footnotemark[1], Huan Zhao\footnotemark[1], Haojie Pan\footnotemark[1], Yangqiu Song\footnotemark[1], Vincent W. Zheng\footnotemark[2], Qiang Yang\footnotemark[2]\footnotemark[1]}
\email{{wxiaoae, hzhaoaf, hpanad, yqsong}@cse.ust.hk;
{vincentz, qiangyang}@webank.com}
\affiliation{\institution{\footnotemark[1]Hong Kong University of Science and Technology, Hong Kong}
\institution{\footnotemark[2]WeBank, China}
}

%
\renewcommand{\shortauthors}{Xiao et al.}

%
\begin{abstract}
An effective content recommendation in modern social media platforms should benefit both creators to bring genuine benefits to them and consumers to help them get really interesting content. In this paper, we propose a model called Social Explorative Attention Network (SEAN) for content recommendation. SEAN uses a personalized content recommendation model to encourage personal interests driven recommendation. Moreover, SEAN allows the personalization factors to attend to users' higher-order friends on the social network to improve the accuracy and diversity of recommendation results. Constructing two datasets from a popular decentralized content distribution platform, Steemit, we compare SEAN with state-of-the-art CF and content based recommendation approaches. Experimental results demonstrate the effectiveness of SEAN in terms of both Gini coefficients for recommendation equality and F1 scores for recommendation performance.
\end{abstract}

%
%


%

%

%
\maketitle

\vspace{-0.2in}
{\fontsize{8pt}{8pt} \selectfont\textbf{ACM Reference Format:}\\
Wenyi Xiao, Huan Zhao, Haojie Pan, Yangqiu Song, Vincent W. Zheng, Qiang Yang. 2019. Beyond Personalization: Social Content Recommendation for Creator Equality and Consumer Satisfaction.
In \textit{The 25th ACM SIGKDD
Conference on Knowledge Discovery \& Data Mining (KDD'19), August 4--8, 2019, Anchorage, AK, USA.} 
ACM, NY, NY, USA, 11 pages.\\ 
https://doi.org/10.1145/3292500.3330965}

\section{Introduction}
\label{sec-intro}

Content recommendation, such as news recommendation,  has been studied for many years in recommender systems (RSs). 
When applying to modern content distribution platforms, such as Facebook and Steemit\footnote{Steemit (https://steemit.com/) is a blockchain based social media and decentralized content distribution platform for consumers and creators to earn Steemit tokens by playing with the platform and interacting with others. It is regarded as a more effective content distribution ecosystem that allows small content creators to share their creative contents while protecting the copyright without any intermediaries. }, an effective content recommendation algorithm should consider both content creators to bring genuine benefits to them and content consumers to help them get really interesting contents.
While more accurate recommendation can improve the consumers' reading experience, it is regarded as a healthier content distribution ecosystem that encourages individuals, especially small content creators, to share their creative contents.

However, existing recommendation algorithms still lack consideration for balancing both content creators and consumers.
Content recommendation methods can be content based or collaborative filtering (CF) based ones.
Content based methods~\cite{wang2017dynamic,wang2018dkn} memorize historical reading/watching content of a user and predict his/her future reading/watching content based on features or similarities of both contents. 
Such approaches emphasize on particular topics for a user and may not be able to encourage diversity of recommendation results unless a content consumer actively searches for new topics.

CF is considered as a complementary technique to content based approaches for content recommendation~\cite{das2007google} as it can oversee the global users' clicking behaviors and make recommendation based on similar users or similar contents, where similar contents mean contents clicked/read by the same group of users.
CF usually optimizes based on global behavior information so that the platform will attract more clicks or reading actions.
Unfortunately, CF will produce unintended Matthew's Effects (``The Rich Get Richer'')~\cite{MatthewEffect} which
will hurt small/new content creators who may not be able to attract attentions.
Although most of the traditional CF based methods are often called personalized recommendation~\cite{das2007google,liu2010personalized} and can be generalized to social networks using social regularization~\cite{jamali2010matrix,ma2011recommender,ye2012exploring,yang2012circle,zhao2017sloma,sun2018attentive,chen2019social}, they are in nature looking at global information and cannot solve this problem.

\begin{figure}
\centering
\includegraphics[width=0.4\textwidth]{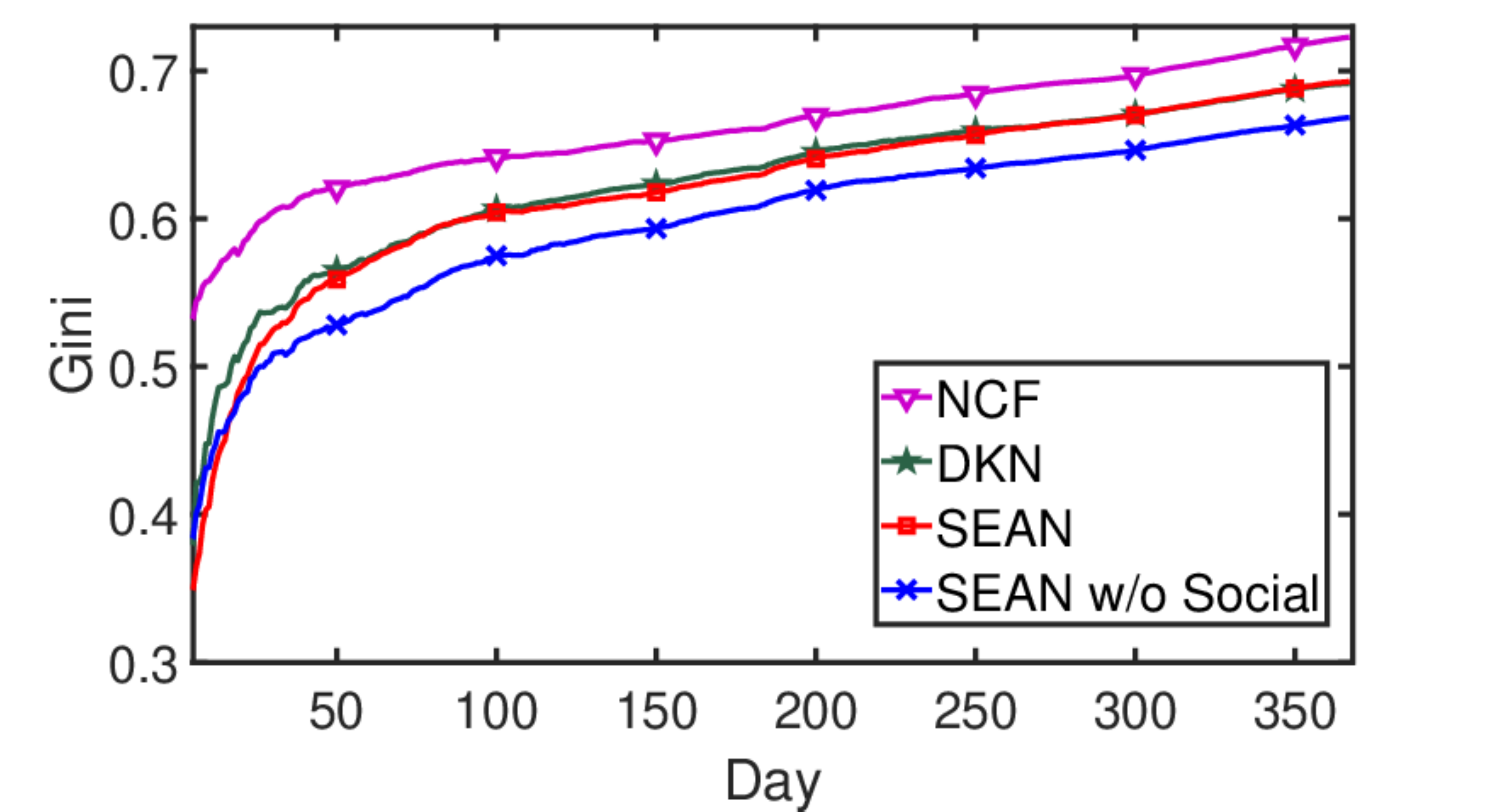}
\vspace{-0.1in}
\caption{\small Comparison of Gini coefficients of different algorithms for 368 days using Steemit social media. 
Gini coefficient is computed over the distribution of recommendation impression numbers of content creators.
We compared content based DKN~\cite{wang2018dkn} (F1=42.85), NCF~\cite{he2017neural} (F1=42.14), and our algorithm SEAN with (F1=47.69) and without any social information (F1=42.40). 
\vspace{-0.3in}
}
\label{gini-comparison}
\end{figure}

There have been a few recommender systems ~\cite{SalganikDodds,AbeliukBHHL17} that have studied the effects related to Matthew's Effect.
However, they are still CF based recommendation and the recommendation strategies are relatively simple, e.g., using popularity~\cite{SalganikDodds} or quality~\cite{AbeliukBHHL17,BerbegliaH17} of the content.
Moreover, one possible way to reduce Matthew's Effect is to use mechanism design approaches in game theory~\cite{BerbegliaH17}. 
In fact, the developed strategy (e.g., introducing randomness in recommendation~\cite{BerbegliaH17}) may hurt consumer's satisfaction as the recommended content may be not related to a consumer's interests, but such effect has not been considered and discussed yet.
As shown in Figure~\ref{gini-comparison}, we use the Gini coefficient over the distribution of recommendation impression numbers for content creators on Steemit social media to demonstrate Matthew's Effect, as the Gini coefficient is usually used to measure inequality and large Gini values mean eventually a small number of content creators dominating the content consumption. 
We also use the F1 score of prediction to evaluate content consumer's satisfaction.
From the Figure~\ref{gini-comparison}, we can see that the state-of-the-art CF based approach NCF~\cite{he2017neural}  encourages inequality more than content based approach DKN~\cite{wang2018dkn} although their F1 scores are comparable to each other.
Thus, a natural question is {\it can we have a content recommendation algorithm that can further benefit content consumers while not hurting creators?}

In this paper, we consider both creators and consumers.
For {\it creator equality}, we also use a content based approach.
However, as a traditional content based approach, DKN encodes the new incoming document into a unique vector (the same vector that will be compared with all users) and uses this vector to attend to a user's historically read documents. 
In this way, popular content will still tend to be selected by the final prediction classifier regardless of the user's personal interests.
Different from DKN, we use user-dependent vectors to attend to related words and sentences in a new incoming document.
In this way, we compress a user's interests into contextual vectors and use the user-dependent document representation vector to feed the final prediction classifier.
This is more compatible with the personalized nature of content recommendation that can benefit small creators, as long as the created content is of the consumer's interests.
As shown in Figure~\ref{gini-comparison}, our model without social information can achieve comparable F1 while significantly reduce the Gini coefficient.
A natural way to further encourage diversity and improve creators equality is to introduce more randomness as suggested by the mechanism design approach~\cite{BerbegliaH17}. We demonstrate this by randomly exploring other consumers' interests. However, this mechanism will hurt the prediction accuracy which reflects the consumer's satisfaction.

To improve the {\it consumer satisfaction},  as the training data for each user's context vectors may be too small and cannot train well for users having less reading history, we allow a user's context vectors to ``attend'' to friends' context vectors and fetch back friends' reading interests and prediction knowledge by aggregating their context vectors.
Nonetheless, even a user has many one-hop friends, friends sharing a similar topic of interest may not be enough.
Therefore, we consider the user's higher-order friends.
An extreme case is that we go over all $n$-hop friends, and it is likely that we can reach all connected users in a social network when $n$ is large.
Apparently, it will be too expensive to explore.
To remedy this, we develop a social exploration mechanism based on Monte Carlo Tree Search (MCTS)~\cite{MasteringGo2016}.
This will be more effective to attend to higher-order friends.
In particular, by using MCTS, we can achieve a good balance of finding $n$-hop friends with similar interests and exploring friends with some randomness for more diverse interests in a social network.
As shown in Figure~\ref{gini-comparison}, introducing social information will significantly improve the F1 score but also increase the Gini coefficient.
Therefore, in the experiments, we systematically study how different hyper-parameters of including social information can affect both prediction accuracy and the Gini coefficient.

Our contributions can be highlighted as follows:

$\bullet$ We consider both content creators and consumers in content recommendation. 
In particular, we use the Gini coefficient to measure the inequality of content creators based on recommendation impressions.

$\bullet$ We propose a novel social explorative attention based recommendation model to use a user's personal reading history and go beyond personal data to explore the user's higher-order friends.

$\bullet$ We construct two datasets of different languages from a popular decentralized content distribution platform, Steemit. By conducting extensive experiments, we demonstrate the superiority of our model over existing state-of-the-art recommendation approaches, including CF and content based ones, in terms of benefiting both consumers and creators.

The code and data are available at https://github.com/HKUST-KnowComp/Social-Explorative-Attention-Networks.

\vspace{-0.05in}

\section{Overview}

We propose to use a personalized model to perform content recommendation, as personalization will encourage the model to find more relevant contents and less affected by the global information about popularity and social influence.
Then we socialize it to make the personalized factors be able to attend to friends' information, which will further balance the randomness factor to improve the creator equality and the relatedness factor to improve the recommendation performance.
We call our recommendation model  Social Explorative Attention Network (SEAN).

For the document representation, we adopt the hierarchical attention networks~\cite{yang2016hierarchical}.
In the word level, the attention is used to select useful words to construct features to feed to sentence representations.
In the sentence level, the attention is used to select useful sentences to construct features for the whole document and then fed to the final classifier.
We use a user-dependent attention model to personalize the document representation learning.
Since each user has different word and sentence level attentions on a document, the document representation based on attentions will be different for different users.

\begin{figure}[t]
\vspace{-0.1in}
\centering
\includegraphics[width=0.36\textwidth]{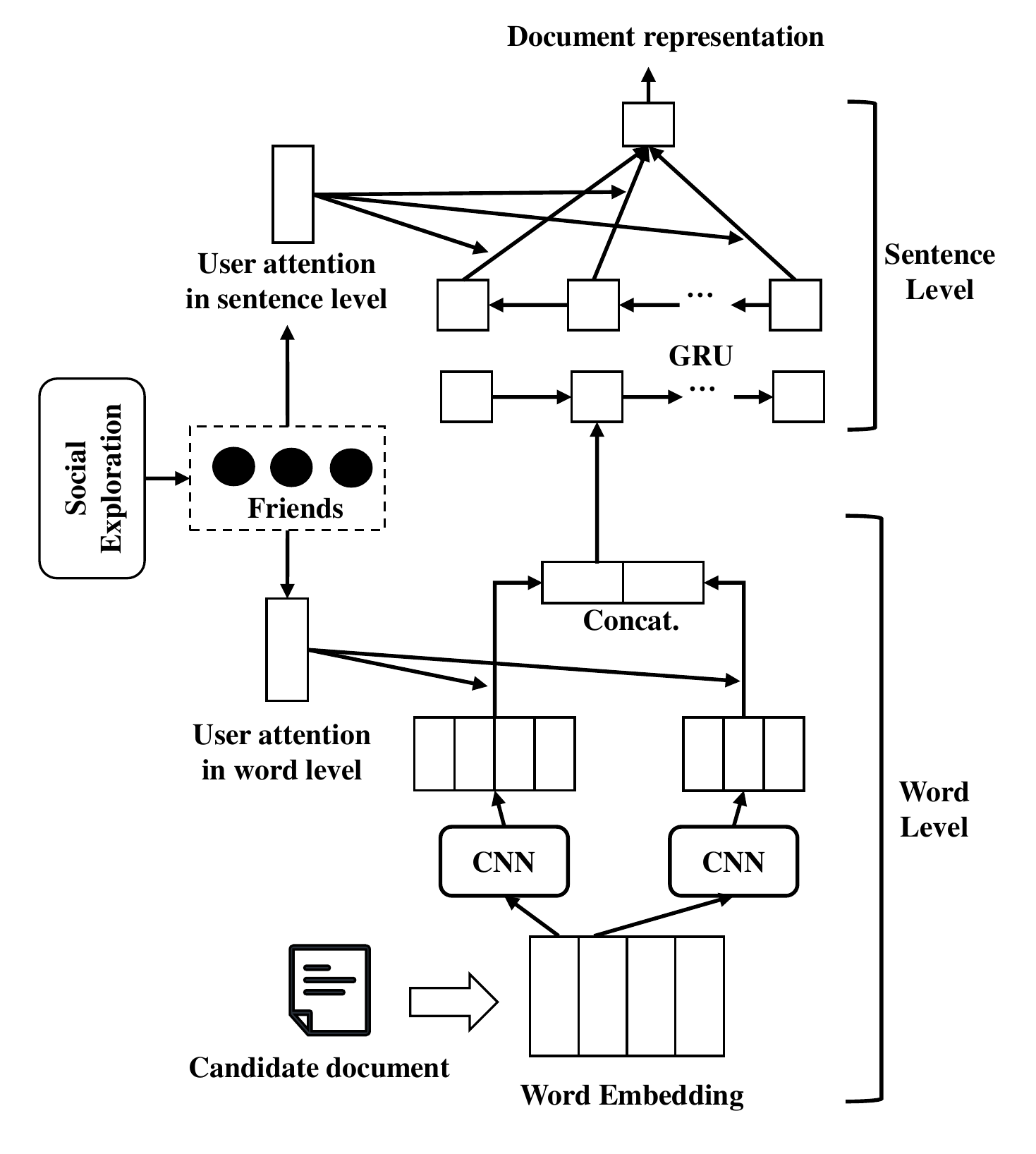}
\vspace{-0.2in}
\caption{\small The architecture of SEAN. The left side is a social exploration module which explores high-order friends for the RS system in the ride side. These friends are incorporated with the user to build the user's representation vector in word and sentence levels, respectively. The right side is a hierarchical architecture from CNN layer to encode words to GRU layer to encode sentences in the document. The user's representation vectors are used to attend to important words and sentences in the candidate document.}
\label{social-attention}
\vspace{-0.2in}
\end{figure}

For the user representation, we construct user representation vectors (word and sentence levels) themselves as attentions to his/her friends' representation vectors, which is essentially an attention over attention model. 
To enable the attention over attention mechanism to use more information, we propose to explore a user's higher-order friends. The overview architecture of our model is shown in Figure~\ref{social-attention}.


\section{SEAN Recommendation Model}
\label{sec-san}
The recommendation task is to predict whether a target user $u$ will click a given document $d$. 
Here we use the textual document recommendation as an example for content recommendation.
We assume that we have a social graph $\mathcal{G} = (\mathcal{U}, \mathcal{E})$. 
Our goal is to learn a prediction function $p = \mathcal{F}\left ( u, d, \theta  \right )$, where $p$ represents the probability that user $u$ will click a given document $d$, and $\theta$ denotes the model parameters of function $\mathcal{F}$.



\subsection{Socialized Document Representation}
\label{sec-single}
Assume that a document has $I$ sentences and each sentence contains $J$ words. $w_{ij}$ represents the $j$-th word in sentence $s_i$ with the indices $i \in \left [ 0, I\right ]$ and  $j\in \left [ 0, J \right ]$.
We use a hierarchical architecture to learn the document representation.

\subsubsection{Word Level Personalization}
We use pre-trained word embeddings for words and fix them during training. The embedding of each word can be calculated as $ \bw_{ij} = \bW_e \be_{w_{ij}}, i\in \left [ 0, I \right ], j\in \left [ 0, J \right ],
$
where $\bW_e$ is the embedding matrix for all words and $\be_{w_{ij}}$ is a one-hot vector to select one word embedding vector $\bw_{ij}$ for $w_{ij}$.
We concatenate all word embeddings in a sentence to form a sentence matrix $\bW_i \in \mathbb{R}^{D \times J}$ for sentence $s_i$, where $D$ is the dimension size of the embedding vector of each word.
We then use a convolutional neural network (CNN) \cite{kim2014convolutional} to represent sentences in the document. Here, we apply a convolution operation on $\bW_i$ with a kernel $\bK_k \in \mathbb{R}^{g\times r \times D}$, $k \in \left [ 0, K \right ]$ among $K$ kernels of width $g$ and filter size $r$ to obtain the feature $\bf^k$:
\begin{eqnarray}
\bf_{ij}^k = \text{relu}\left ( \bW_i\left [ \ast ,j:j+g-1 \right ]  \odot \bK_k + \bb_k \right),
\end{eqnarray}
where $j \in \left [ 1, J-g+1\right ]$ is the iteration index of convolution, $\bf_{ij}^k \in \mathbb{R}^{r}$ is regarded as $j$-th CNN feature by the $k$-th kernel $K_k$, and the bias vector $\bb_k$ in $i$-th sentence. 


We then feed each CNN features $\bf_{ij}^k$ to a non-linear layer, parameterized by a global weight matrix $\bW_w \in \mathbb{R}^{{h}\times r}$ to get a hidden representation ${\bf_{ij}^k}'$. $h$ and $r$ are pre-defined dimensions of hidden vectors. 
We measure the importance of the word towards the target user as the similarity of ${\bf_{ij}^k}'$ and the word-level user's socialized representation vector $\bx_w$ (which will be introduced in Section \ref{sec-user}). The sentence representation vector $\bs_i^k$ by CNN with kernel size $k$ is computed as a weighted sum based on the soft attention weights:
\begin{eqnarray}
&& {\bf_{ij}^k}' =\tanh(\bW_w \bf_{ij}^k+\bb_w), \\
&& \alpha_{ij} = \text{Softmax}(\bx_w^{\top} {\bf_{ij}^k}'),\\
&& \bs_i^k=\sum_{j}\alpha_{ij} \bf_{ij}^k, 
\end{eqnarray}
where the superscript $\cdot^\top$ represents the vector or matrix transpose. All representation vector $\bs_i^k$ are concatenated together and taken as the sentence embedding $\bs_{i}$ for sentence $s_i$ as:
$
\bs_{i} = \left [ \bs_i^1, \bs_i^2, ..., \bs_i^K\right].
$

\subsubsection{Sentence Level Personalization}
At the sentence level, we use a bidirectional Gated Recurrent Unit network (Bi\-GRU) \cite{bahdanau2014neural} to compose a sequence of sentence vectors into a document vector. The BiGRU encodes the sentences from two directions:
\begin{equation}
    \bh_{i}  =   \overrightarrow{GRU} (\bs_{i}) \left |  \right |  \overleftarrow{GRU} (\bs_{i}).
\end{equation}
After getting $\bh_{i}$ for sentence $s_i$, we use the sentence level user representation vector $\bx_s$ to extract relevant sentences that are interested by the target user and get a final document representation $\bd$ by soft attention mechanism similar to sentence representation. We omit the details of equations due to the lack of space and the similarity with the word level computation.
As shown in the right side of Figure~\ref{social-attention}, we have two layers of feature extraction networks.
This architecture is inspired by~\cite{yang2016hierarchical} since it is better for long document modeling.
In our model, we use CNN instead of RNN for word level since in practice we found that CNN is faster, more robust, and less easy to overfitting on our datasets.
Moreover, different from \cite{yang2016hierarchical}, we use socialized user representation vectors instead of unified representation vectors for attending to words and sentences.

\subsection{Socialized User Representation}
\label{sec-user}

We denote $\be_u$ as a one-hot vector of user $u$ and retrieve the word level user representation $\bu_w$ from a trainable embedding matrix $\bA\in\mathbb{R}^{h\times |\mathcal{U}|}$ by using $\bA\be_u$, where $h$ is the size of attention vectors.
We can get the user's sentence-level representation by another trainable embedding matrix $\bA'$ in the same way.
%
We design a social attention module to enrich a user's representation by incorporating his/her friends' representations. 
Let $\by_i\in \mathbb{R}^{h}, i \in \left \{ 1,2,... \right \}$ be his/her friends' word-level representation vectors, and denote $\by_0 = u_w$. 
The attention mechanism produces a representation $\bx_w$ as a weighted sum of the representations vectors $ \by_j, j \in \left \{ 0,1,2,... \right \}$ via
\begin{eqnarray}
&& \alpha_j = \text{Softmax}(\text{LeakyReLU}(\bw^\top\left [ \bW_y \bu_w || \bW_y \by_j\right ])),\\
&& \bx_w = \sum_{j}\alpha_{j} \bW_y \by_j,
\end{eqnarray}
where $\bW_y \in \mathbb{R}^{h \times h}$ is a shared linear transformation and $||$ is the concatenation operation. The attention mechanism is a single-layer feedforward neural network, parametrized by a weight vector $\bw \in \mathbb{R}^{2h}$, and applying the LeakyReLU nonlinearity.

Similarly, we can get the sentence-level user representation $\bx_s$ by the attention of high-order friends' representation vectors.


\subsection{Prediction and Learning}

Finally, we use a dense layer to predict the probability that the target user $u$ will read the candidate document $d$:
\begin{equation}
p =\mathrm{Softmax} (\bw_g^\top \bd + \bb),
\end{equation}
where $\bw_g \in \mathbb{R}^{2h}$ is a global trainable weight vector trained by all the samples. $\bd$ is the document representation vector obtained from Section \ref{sec-single}.

Due to the nature of the implicit feedback and the task of item recommendation, we adopt the binary cross-entropy loss to train our model:
\begin{equation}
    \mathcal{L}(\theta)= -\frac{1}{M}\sum_{m=1}^{M} \left[y_m \log(p_m) + (1-y_m)\log(1-p_m) \right],
\end{equation}
where $m$ is the index of a sample, $M$ is the total number of training samples, $y_m\in \{0,1\}$ is the  label, and $\theta$ denotes the set of model parameters. The negative samples are formed from the documents that the target user does not make response to while his/her friends make. 

During training and testing, we train the model with the data of past $t$ days and test it with the data on $(t+1)$-th day. The model dynamically adapts to new data day-by-day.

\begin{figure*}[t]
\centering
\includegraphics[width=0.8\textwidth]{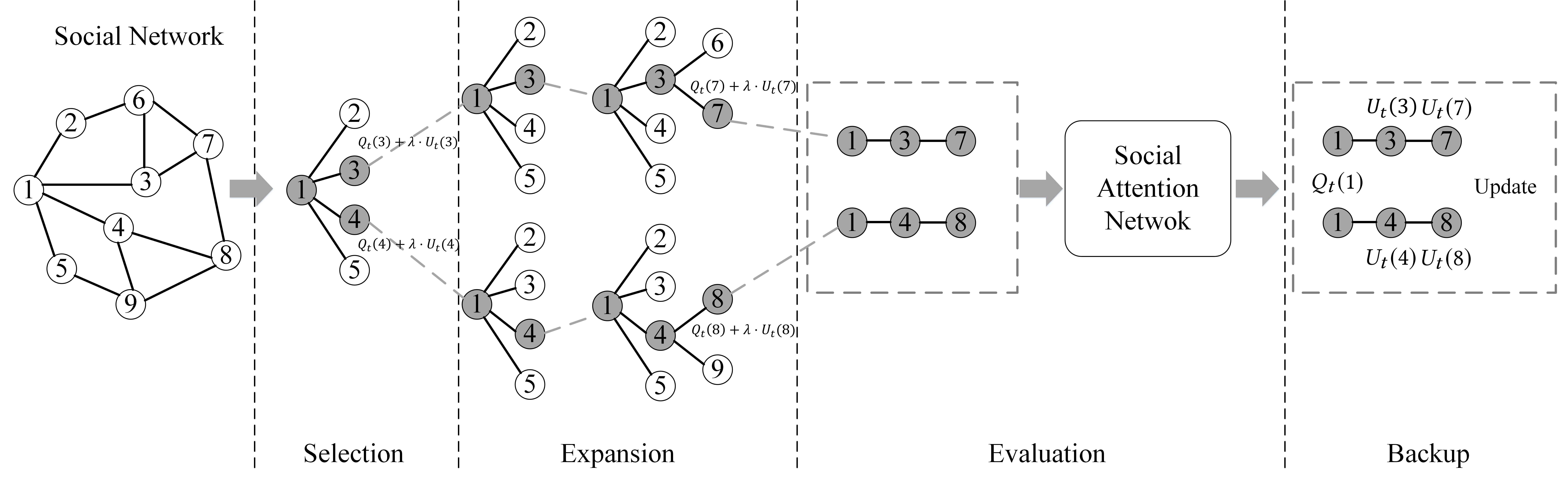}
\caption{\small MCTS for social exploration. We illustrate our MCTS based strategy with this example. We set the beam width and search depth to 2. The node "1" represents the target user. We initialize 2 paths according to the beam width and add "1" to each path. In the Selection step, we calculate the scores by Eq. (\ref{eq-ucb1}) of "1"'s neighbors ("2", "3", "4", "5") and select two nodes with the largest scores ("3", "4") and add them to each path. In the Expansion step,  we calculate the scores for the neighbors of both "3" and "4", and again select two nodes ("7", "8") among all the neighbors. In the Evaluation step, the two generated path ("1"->"3"->"7" \& "1"->"4"->"8") are input to RS model mentioned in Section \ref{sec-san}. In the same way, we get the paths for other nodes. In the Backup step, we update $Q_t(1)$ from the result at Evaluation and $U_t(3), U_t(4), U_t(7), U_t(8)$ from Selection \& Expansion.}
\label{fig-mcts}
\vspace{-0.1in}
\end{figure*}

\subsection{Complexity Analysis}
\label{sec-complexity}
For the word level, the time complexity is linear to the number of tokens in the training data set, which is $M \cdot I \cdot J$, where $M$ is the number of training samples,  $I$ is the maximum number of sentences, and $J$ is the maximum number of tokens in a sentence.
It is also linear to the number of kernels $K$, the numbers of hidden vectors $h$, the filter size $r$, and the convolutional window size $g$. Since we use fixed-sized word embeddings, the large number of words do not contribute to our time cost.
For the sentence level, the time cost of the GRU layer is linear to the maximum number of sentences $I$ and the number of parameters in the GRU cell $P_{GRU}$.
For both word-level and sentence-level attentions, the cost is linear to the square number of hidden dimension $h^2$, the number of selected friends $L$, and the times of trials of attention $B$. Note that the number of selected friends $L$ and the times of trials of attention $B$ are the same to the search depth $L$ and beam width $B$  introduced in Section~\ref{sec-se}.
Moreover, for the fully connected layer, the parameter is linear to $h$.
Therefore, the overall time complexity is 
$O(M \cdot I \cdot J \cdot (K\cdot g \cdot r \cdot D  + h \cdot r \cdot K + h^2\cdot L \cdot B) + M \cdot I \cdot (P_{GRU} + h^2\cdot L \cdot B) )$.

\section{Social Exploration}
\label{sec-se}
In this section, we first introduce how we use MCTS for friends selection, and then further enhance MCTS with beam search.

\subsection{Selecting Friends with MCTS}
\label{sec-ucb}

Monte Carlo Tree Search (MCTS)~\cite{MasteringGo2016} is a stochastic search algorithm to find an optimal solution in the decision space.
It models an agent that simultaneously attempts to acquire new knowledge (called ``exploration'') and optimize the decisions based on existing knowledge (called ``exploitation''). 
MCTS uses the upper confidence bounds one (UCB1)~\cite{kocsis2006bandit} value to determine the next move $a$ from a viewpoint of multi-armed bandit problem. The selection strategy is defined by: 
\begin{equation}
\label{eq-ucb1}
a = \argmax_v\{ Q_t(v) + \lambda \cdot U_t(v) \},
\end{equation}
where $Q_t(v)$ denotes the empirical mean exploitation reward of node $v$ at time $t$ and $U_t(v)$ is the utility to explore node $v$. This equation clearly expresses the exploration-exploitation trade-off: while the first term of the sum tends to exploit the seemingly optimal arm, the second term of the sum tends to explore less pulled arms. $\lambda$ is used to balance the two terms.

Here we explain how MCTS guides to generate a path with a fixed number of search depth $L$, regarded as $L$ friends of $u$ by walking through the social graph. We denote the target user $u$ as the starting node $c_0$ and denote $c_l, l\in \left [ 0, L \right ]$ as the $l$-th node added for $u$ in the path.
On day $t$ at search step $l$, the node $c_{l+1}$ is retrieved from the neighbors of node $c_l$. We calculate the score for each neighbor according to Eq.~\eqref{eq-ucb1} and choose the neighbor with the maximum score as the $(l+1)$-th friend of the user $u$.  The design of calculating the values from exploitation and exploration are mentioned below.

\subsubsection{Exploitation}
\label{sec-mcts-exploitation}
On day $t$, we compute the $Q_t(v)$ to get the exploitation reward of neighbor node $v$. In our scenarios,
we want to select those as friends who can improve the recommending performance as much as they can. In this work, we design four exploitation strategies to select friends for maximizing $Q_t(v)$: the average F1 from RS model (SEAN-RS-F1), static PageRank value from social network (SEAN-SPR), dynamic PageRank value from activity network (SEAN-DPR), as well as the actual payout each user earned in blockchain platforms (SEAN-Payout).

\textbf{SEAN-RS-F1.}
We regard the average F1 evaluated based on our RS model of each neighbor node $v$ up to time $t$ as the exploitation reward $Q_t(v)$. This is based on the assumption that a user who has been well-learned by the RS model is reliable and could be exploited as a friend for the target user $u$ in the future. In this way, the RS prediction results can guide the friend exploration process, and in turn, the friend exploration process provides useful friends to help enrich the target user's representation. 

\textbf{SEAN-SPR.}
The second way is to use the PageRank value of $v$ obtained from social network as exploitation reward $Q_t(v)$. In \cite{xiang2013pagerank}, Xiang et.al. explicitly connect PageRank with social influence model and show that authority is equivalent to influence under their framework. Thus, we assume that a node with high PageRank value in the social network is influential and should be exploited as a friend for the target user $u$. 

\textbf{SEAN-DPR.}
On social media platforms, each user can not only make activities on the documents (as a consumer) but also create documents (as creator). We build a dynamic activity network and calculate the PageRank values of nodes. Compared with the social network, the edges in the activity network are the consumer-creator connection. 

\textbf{SEAN-Payout.}
In some blockchain based social platforms, the platform would give some rewards, i.e., bitcoin, to those users who help distribute the documents, i.e., post or forward a document in the platform. 
We regard the payout that a user gains as the value of his/her exploitation value $Q_t(v)$.

\subsubsection{Exploration}
We design the exploration mechanism to get the explored reward $U_t(v)$ for friend $v$ as follows:
\begin{equation}
U_t(v) = \sqrt{\frac{\text{log} N_t(c_l)}{N_t(v)+ 1}},
\label{eq-ucb-u}
\end{equation}
where $N_t(c_l)$, $N_t(v)$ denote as the times that the current node $c_l$ at search step $l$ and the neighbor node $v$ have been selected as friends up to day $t$, respectively. The goal of the exploration is to select the nodes who have less been explored in the past.

\subsection{Obtaining Multi-paths with Beam Search}
If we want to find higher-order friends, we can greedily select the next node with a maximum score from the neighbors of the current node at search step  $l$. In this way, we would get a path of higher-order friends. If we want to find more than one paths, it is time-consuming to get a globally optimal set of paths. Therefore, we combine MCTS with beam search~\cite{koehn2004pharaoh} to balance the optimality and completeness. At search step $l$, we choose the neighbors with largest $B$ scores from Eq.~(\ref{eq-ucb1}) and these $B$ nodes are selected for further expansion. Here $B$ is the beam width. In this way, we generate $B$ paths for the target user $u$. For training and testing, we obtain $B$ prediction results by using each path and $u$ and compute the average of these results to get the final prediction. We give a concrete example of MCTS for social exploration, shown in Figure \ref{fig-mcts}. We introduce how we select friends based on beam MCTS in Algorithm~\ref{alg-bmcts} and how it is used to train SEAN in Algorithm~\ref{alg-sean} in Appendix~\ref{sec:appendexi-bmcts}.





\section{Experiments}
\label{sec-exp}

\subsection{Dataset Description}
\label{sec-dataset}
We build two datasets, Steemit-English and Steemit-Spanish from the decentralized social platform, Steemit.
Steemit is a blogging and social networking platform that uses the Steem blockchain to reward creators and consumers. 
Most of the modern content distribution platforms are already using recommendation systems to recommend contents to users, which can be biased if we collected data from them for our evaluation. 
Different from them, the contents and user clicks are not manipulated by the Steemit platform.
We retrieve the commenting activities of users (consumers) from June \nth{2}, 2017 to July \nth{6}, 2018. 
Two datasets are constructed based on social communities using English and Spanish respectively.

We form a sample as a triplet with three elements: a given user, a document, and a label $1/0$. We form the positive samples by the documents which users have made comments. We treat messages that users' friends have made responses but the users themselves do not as negative samples.  Since we collect users' activities information from their comment logs, it is natural that the number of users who made comments on this platform is not too much. 
The statistics of the two datasets are shown in Table \ref{table-data-descript}. 

\begin{table}[t]
    \centering
    \caption{Statistics of the two datasets.}
    \label{table-data-descript}
    \vspace{-0.1in}
    {\footnotesize
\begin{tabular}{c|c|c}
\toprule
 & \textbf{Steemit-English} & \textbf{Steemit-Spanish} \\

\midrule
Duration (days)  & 370  & 126 \\
\# Consumers & 7,242 & 1,396 \\
\# Creators & 44,220 & 4,073 \\
\# Relations & 273,942 & 25,593 \\
\# Documents & 177,134 & 14,843 \\
Avg. word per document & 290 & 509\\
\# Logs & 220,909 & 20,893\\
\# Samples & 684,752 & 92,236\\

\bottomrule
\end{tabular}
}
\vspace{-0.2in}
\end{table}

\subsection{Evaluation Metrics}
\label{sec-metrics}
To evaluate the recommendation quality of the proposed approach, we use the following metrics: Area under the Curve of ROC (AUC) and F1 for consumer satisfaction and the Gini coefficient for creator equality, where Gini coefficient is defined as:
$
    \text{Gini} = \frac{\sum_{i=1}^{n}(2i-n-1)x_i}{n\sum_{i=1}^{n}x_i},
$
$x$ is an observed value, $n$ is the number of values observed, and $i$ is the rank of values in ascending order.
To measure the performance of models considering both creators and consumers, we calculate the harmonic mean of F1 and (1-Gini), denoted as C\&C:
\begin{equation}
    \text{C\&C} = \frac{2\times (1-\text{Gini})\times \text{F1}}{(1-\text{Gini})+\text{F1}}.
\end{equation}
Since we train and test day-by-day, we compare a model's quality by the average of each metric during the whole peorid. For AUC, F1, and C\&C, the larger, the better. For Gini, the smaller, the better.

\subsection{Baselines}
\label{sec-baselines}
We compare our model with following baselines.

\textbf{LR} \cite{van2013deep} is the simplest word-based model for CTR prediction. We use TF-IDF to extract keywords for a user's clicked historical documents and the new incoming document and feed them to a logistic regression model to predict the label. 

\textbf{LibFM} \cite{rendle2012factorization} is a state-of-the-art feature-based factorization model and widely used in CTR prediction. In this paper, we use the same features as LR and feed them to LibFM. LibFM treats a user's features and a document's features separately for the factorization.


\textbf{DKN}~\cite{wang2018dkn} learns representations of documents and users. In DKN, it obtains a set of embedding vectors for a user's clicked historical documents. 

Then an attention is applied to automatically match the candidate document to each piece of his/her clicked documents, and aggregate them with different weights. 
Here, we only use DKN's base model without the knowledge graph information.

\textbf{NCF} \cite{he2017neural} is short for Neural network based Collaborative Filtering. It is a deep model for recommender systems which uses a multi-layer perceptron (MLP) to learn the user$-$item interaction function. It ignores the content of news and uses the comment counting information as input.

\textbf{SAMN}~\cite{chen2019social}, Social Attentional Memory Network, is a collaborative filtering model which employs the attention-based memory module to attend to a user's one-hop friends' vectors. The attention adaptively measures the social influence strength among friends. 

\textbf{SEAN}, if without any clarification, is SEAN-RS-F1, which is our model using F1 score as the exploitation value.

\subsection{Experimental Settings}
\label{sec-setup}
For our framework, we use pre-trained word embeddings for the document and fix them during training.
For the word-level representation in the CNN layer, the filter number is set as 50 for each of the window sizes ranging from 1 to 6.
The hidden vector size is set to 128 for both GRU layers and dense layers. 
The beam width $B$ is set to 3 and $\lambda$ is set to 1. 
The search depth $L$ is set to 10.
We train the model for 6 epochs every day. The data in every day is split to 9:1 for training and validation.
We train the model from the data of past $t$ days and test it by using the data on $t+1$-th day. More experimental settings are shown in Appendix~\ref{sec:appendix-settings}.

The key parameter settings for baselines are as follows. 
For the keyword extraction in LR and LibFM, we set the number of keywords for document and user's historical readings as 20 and 90. 
For DKN, the length of the document embedding is set to 200 and due to the limitation of memory, and we use a user's latest 10 clicked documents to represent the user.
The above settings are for fair consideration. Each experiment is repeated five times, and we report the average and standard deviation as results.

\begin{table*}[t]
    \centering
    \caption{
    Comparison of different methods on Steemit datasets. The best results are highlighted in boldface.
    }
    \label{tb-performance}
    \vspace{-0.1in}
{\footnotesize
\begin{tabular}{c|c|c|c|c|c|c|c|c}
\toprule
\multirow{2}{*}{} & \multicolumn{4}{c|}{\textbf{Steemit-English}} & \multicolumn{4}{c}{\textbf{Steemit-Spanish}} \\ \cline{2-9}
 & AUC & F1 & Gini & C\&C & AUC & F1 & Gini & C\&C \\ \midrule
NCF & 52.83$\pm$0.13 & 42.14$\pm$0.21 & 66.04$\pm$0.25 & 37.71$\pm$0.22 & 50.46$\pm$0.21 & 35.02$\pm$0.26 & 58.13$\pm$0.34 & 38.14$\pm$0.29 \\ 
SAMN & 53.05$\pm$0.35 & 42.28$\pm$0.45 & 65.98$\pm$0.21 & 37.80$\pm$0.28 & 51.10$\pm$0.24 & 35.24$\pm$0.31 & 58.29$\pm$0.32 & 38.20$\pm$0.31 \\ 
LR & 52.89$\pm$0.07 & 34.50$\pm$0.11 & 62.89$\pm$0.11 & 35.86$\pm$0.11 & 53.15$\pm$0.06 & 36.50$\pm$0.29 & 55.84$\pm$0.09 & 39.97$\pm$0.14 \\
LibFM & 50.01$\pm$0.12 & 40.43$\pm$0.22 & 66.42$\pm$0.13 & 36.79$\pm$0.16 & 47.71$\pm$0.30 & 22.37$\pm$0.33 & 56.50$\pm$0.21 & 29.55$\pm$0.26 \\ 
DKN & 62.71$\pm$0.22 & 42.85$\pm$0.45 & 62.29$\pm$0.26 & 40.22$\pm$0.33 & 57.02$\pm$0.39 & 41.27$\pm$0.45 & \textbf{53.98$\pm$0.25} & 43.52$\pm$0.32 \\ \hline
\textbf{SEAN} & \textbf{65.57$\pm$0.17} & \textbf{47.69$\pm$0.46} & \textbf{61.78$\pm$0.24} & \textbf{42.43$\pm$0.33} & \textbf{59.98$\pm$0.34} & \textbf{42.99$\pm$0.37} & 53.99$\pm$0.23 & \textbf{44.46$\pm$0.28} \\
\bottomrule
\end{tabular}
}
    \vspace{-0.1in}
\end{table*}

\subsection{Results and Analysis}
\label{sec-results}

Table \ref{tb-performance} reports the results on Steemit-English and Steemit-Spanish datasets.  For consumers, SEAN improves F1 by above 5 percentage point and AUC by near 3 percentage point compared with the best content-based model DKN on Steemit-English and improves F1 by 1.7 percentage point and AUC by near 3 percentage point on Steemit-Spanish. 
This proves that our model can best consider consumer's personal interests and recommend the most interesting contents to them. 
LR and LibFM perform much worse because these two models ignore the word order information and consequently generate worse document and user representations. 
Moreover, compared with CF-based models (NCF and SAMN), SEAN can also outperform them significantly. This result shows that CF methods cannot work well in this recommendation scenario since the documents on Steemit is highly time-sensitive, and the content should be considered for the recommendation.
Besides, from the comparison with the SAMN, we can see that our strategy to incorporate social information is more effective than SAMN. 
We also present a one-year F1 performance of SEAN and DKN on Steemit-English in Figure \ref{baseline-comparison} in Appendix \ref{sec:appendix-overtime}. It further shows that our model has a better performance in most of the days.

For creators, the Gini coefficients of the content-based models are smaller than those of the CF-based models. The result proves our aforementioned claim that CF methods are more likely to suffer from Matthew's effect since CF-based models intend to use global behavioral information to promote popular documents on the social platform. 
The Gini coefficients of SEAN and DKN are comparable, which shows that under the premise of the quality of recommendation for consumers, our algorithm can also encourage creator's equality which may further encourage creators to stay on the platform to keep publishing their innovative contents.

From the harmonic mean C\&C results, we observe that SEAN performs best on both datasets. 
This demonstrates that the social exploration mechanism can have a good balance on optimizing between consumer satisfaction and creator equality.

\subsection{Different Strategies in Social Exploration }
\label{sec-high-oder}

\begin{table}[t]
    \centering
    \caption{Comparison of social exploration methods.}
    \label{tb-exploration-methods}
\vspace{-0.1in}
{\footnotesize
\begin{tabular}{c|c|c|c}
\toprule
Models & F1 & Gini & C\&C \\ 
\midrule
Random Select & 42.48$\pm$0.38 & \textbf{59.13$\pm$0.22} & 41.09$\pm$0.28 \\
Random Walk & 45.05$\pm$0.39 & 60.98$\pm$0.09 & 41.77$\pm$0.20 \\ \hline
SEAN-RS-F1 & \textbf{47.69$\pm$0.46} & 61.78$\pm$0.24 & 42.43$\pm$0.33 \\
SEAN-SPR & 45.99$\pm$0.35 & 60.90$\pm$0.32 & 42.27$\pm$0.33 \\
SEAN-DPR & 45.96$\pm$0.44 & 60.98$\pm$0.22 & 42.21$\pm$0.29 \\
SEAN-Payout & 46.26$\pm$0.36 & 60.65$\pm$0.40 & \textbf{42.53$\pm$0.37} \\ 
\bottomrule
\end{tabular}
}
\vspace{-0.1in}
\end{table}

In the experiment, we evaluate the performance of each exploitation-exploration method using Steemit-English dataset.
``Random Select'' is the model which randomly selects a set of users on the social platform as the target user's friends. 
``Random Walk'' is the model which uses a stochastic process, moving from a node to another adjacent node. These two models are both using random based strategy to explore.
As shown in Table \ref{tb-exploration-methods}, MCTS based models have better F1 than random based models, because the exploitation mechanism can help the model find more relevant friends. SEAN-RS-F1 performs the best on F1 because this model tends to explore friends of higher quality continuously by directly using the recommendation feedbacks. The F1 performance of SEAN-SPR and SEAN-DPR are compatible, while both are worse than the others since SEAN-SPR uses the static social network and SEAN-DPR only uses the daily comment network formed by consumer-creator connections, both missing some information. 
Moreover, MCTS based models also outperform random based models on C\&C, which indicates that our model can improve the recommendation quality for consumers even though slightly hurts the equality. Specifically, SEAN-Payout has the highest C\&C which indicates that using payout, the rewards given by Steemit as the exploitation value, is more suitable to select friends on this platform. Meanwhile, it further verifies the decentralized nature of this platform.      


\subsection{Model Ablation Study}
\label{sec-ablation}

We further compare variants of our SEAN model in terms of following aspects to demonstrate the efficacy of the framework design: the use of social connections, the use of social attention, the use of friend exploration, and different components in the hierarchical document representation. The results are shown in Table \ref{tb-ablation}.

For the consumer side, we can conclude as follows.

$\bullet$ Without any social information means that we are using a pure personalization model for each user. This will decrease F1 by 5 percentage point. This confirms the efficacy of using social information in the SEAN model.

$\bullet$ We also replace social attention with simple averaging friends' representation vectors. This results in a loss of F1 by near 3 percentage point. In other words, it demonstrates the effectiveness of weighing different social influences from friends on recommendation performance.

$\bullet$ We use each user's first-order (one-hop) friends for socialization. 
This is also worse than SEAN with exploring high-order friends, which proves the importance of exploring friends for recommendation.

$\bullet$ We test how CNN for word-level and GRU for sentence-level encoding affect the performance. The usage of CNN and GRU brings about 2 percentage point gain on F1 respectively. Without using GRU and CNN decreases F1 by more than 3 percentage point. 

For the creator side, the Gini coefficient of SEAN w/o social is the lowest one, followed by SEAN with one-hop friends. The reason is that without using any social information, users are not influenced by other users' reading histories, thus cutting off the spread of popular documents. Besides, using the first-order connections is worse than using high-order social information. For models without GRU and/or CNN components, Gini drops within 2 percentage point while F1 also drops. The best result of C\&C indicates that our model can obviously improve consumers' satisfaction without hurting equality too much.

\begin{table}[t]
    \centering
    \caption{Comparison of different variants on SEAN.}
    \vspace{-0.1in}
    \label{tb-ablation}
{\footnotesize
\begin{tabular}{c|c|c|c}
\toprule
Variants & F1 & Gini & C\&C \\
\midrule
w/o social & 42.40$\pm$0.30 & \textbf{58.56$\pm$0.43} & 41.91$\pm$0.35 \\
w/o social attention & 44.79$\pm$0.17 & 62.22$\pm$0.36 & 41.98$\pm$0.23 \\
one-hop friends & 43.08$\pm$0.16 & 60.85$\pm$0.25 & 41.04$\pm$0.20 \\
w/o CNN  & 45.25$\pm$0.22 & 59.98$\pm$0.26 & 42.58$\pm$0.21 \\
w/o GRU  & 45.07$\pm$0.31 & 60.47$\pm$0.14 & 42.12$\pm$0.27 \\
w/o CNN \& GRU  & 44.08$\pm$0.26 & 60.06$\pm$0.20 & 41.91$\pm$0.25 \\
\hline
SEAN & \textbf{47.69$\pm$0.46} & 61.78$\pm$0.24 & \textbf{42.43$\pm$0.33}\\
\bottomrule
\end{tabular}
}
\vspace{-0.2in}
\end{table}

\begin{figure*}[t]
    \subfigure[Search depth $L$.]{\includegraphics[width=1\textwidth]{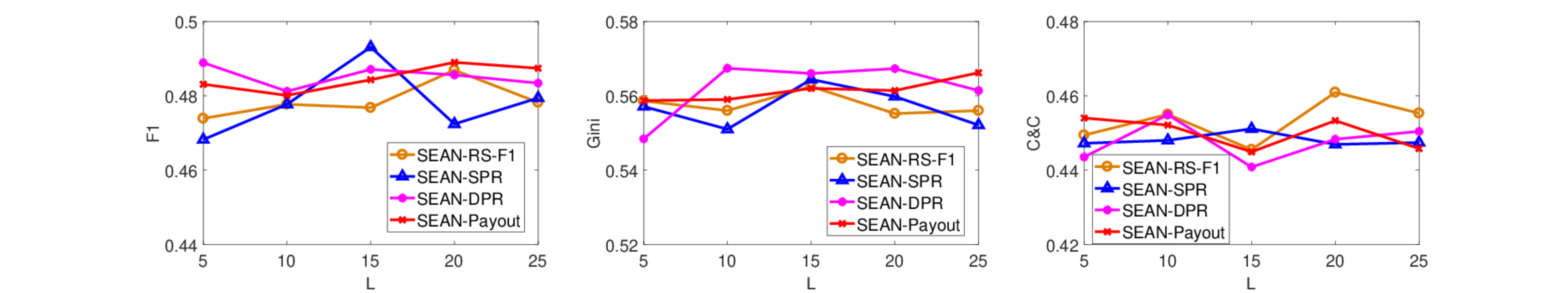}\label{fig-ps-a}}
\\
    \vspace{-0.1in}
    \subfigure[Beam width $B$.]{\includegraphics[width=1\textwidth]{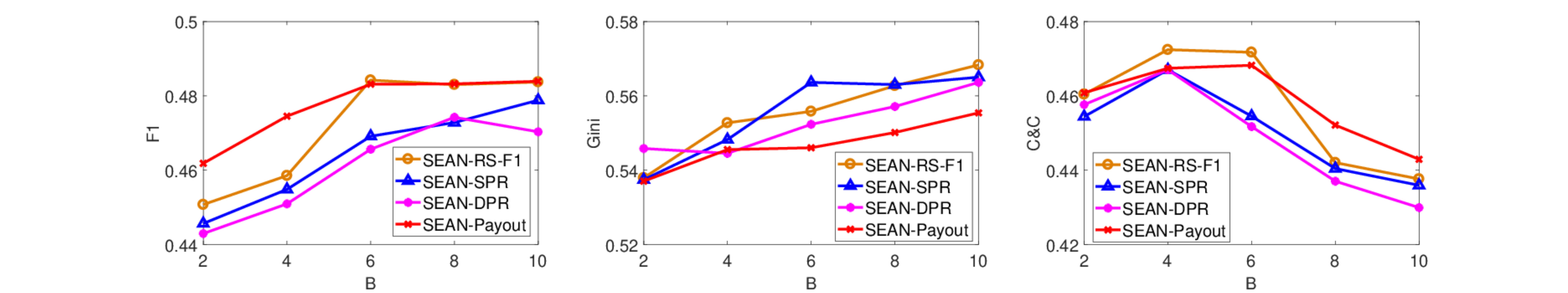}\label{fig-ps-c}}
 \\
    \vspace{-0.1in}
    \subfigure[Trade-off $\lambda$.]{\includegraphics[width=1\textwidth]{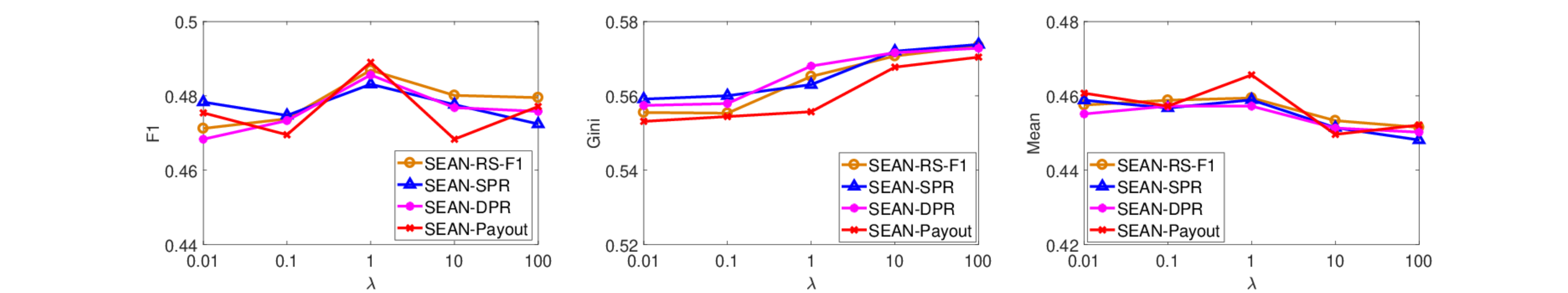}\label{fig-ps-b}}
    \vspace{-0.1in}
    \caption{Hyper-parameter sensitivity analysis using the samples from Steemit-English during the first 100 days.}
    \vspace{-0.1in}
    \label{fig-param}
\end{figure*}

\subsection{Hyper-parameter Sensitivity}
\label{sec-param}
SEAN involves a number of hyper-parameters. Here, we evaluate how different choices of hyper-parameters in social exploration affect the performance of SEAN. In the following experiments, except for the parameter being tested, all other parameters are set as introduced in Appendix \ref{app-sec-setup} if we do not point out. The parameter sensitivity is done by using the samples from Steemit-English during the first 100 days. 

\textbf{ Search Depth $L$.}
We test the influence of search depth $L$ for four proposed models with $L= 5, 10, 15, 20, 25$. The results are shown in Figure \ref{fig-ps-a}. Given the best settings shown in Appendix \ref{app-sec-setup}, changing $L$ from 5 to 25 does not affect both F1 and Gini a lot compared to the beam width $B$. This may indicate that given the Steemit network and the prediction F1 score, using a small number of friends can already cover most of the friends to explore while increasing $B$ will force the exploration to find more neighbors.


\textbf{ Beam width $B$.}
We investigate the influence of the beam width $B$ (number of paths) by setting $B$ ranging from 2 to 10. The results are shown in Figure \ref{fig-ps-c}. 
We can see that F1 increases as the beam width grows since there are more selected friends that are helpful for the user. While with continuing increasing of $B$, F1 tends to be flat since the overlapping of friends selected from each path also increases. Meanwhile, Gini continuously increases when the beam width increases and thus C\&C appears to be best only when $B$ is 4. 

\textbf{ Trade-off constant  $\lambda$.}
The choice of the trade-off constant $\lambda$ is set to be $\lambda\in\left \{0.01, 0.1, 1, 10, 100\right \}$. We can see in Figure \ref{fig-ps-b}, the best F1 is at $\lambda = 1$ for all approaches. It indicates that both exploration and exploitation are important to better select friends. Besides, Gini is less influenced by $\lambda$.

More hyper-parameter evaluations are shown in Appendix~\ref{sec:appendix-parameter}.


\begin{figure}[t]
    \centering
    \subfigure[Search Depth $L$.]{\includegraphics[width=0.23\textwidth]{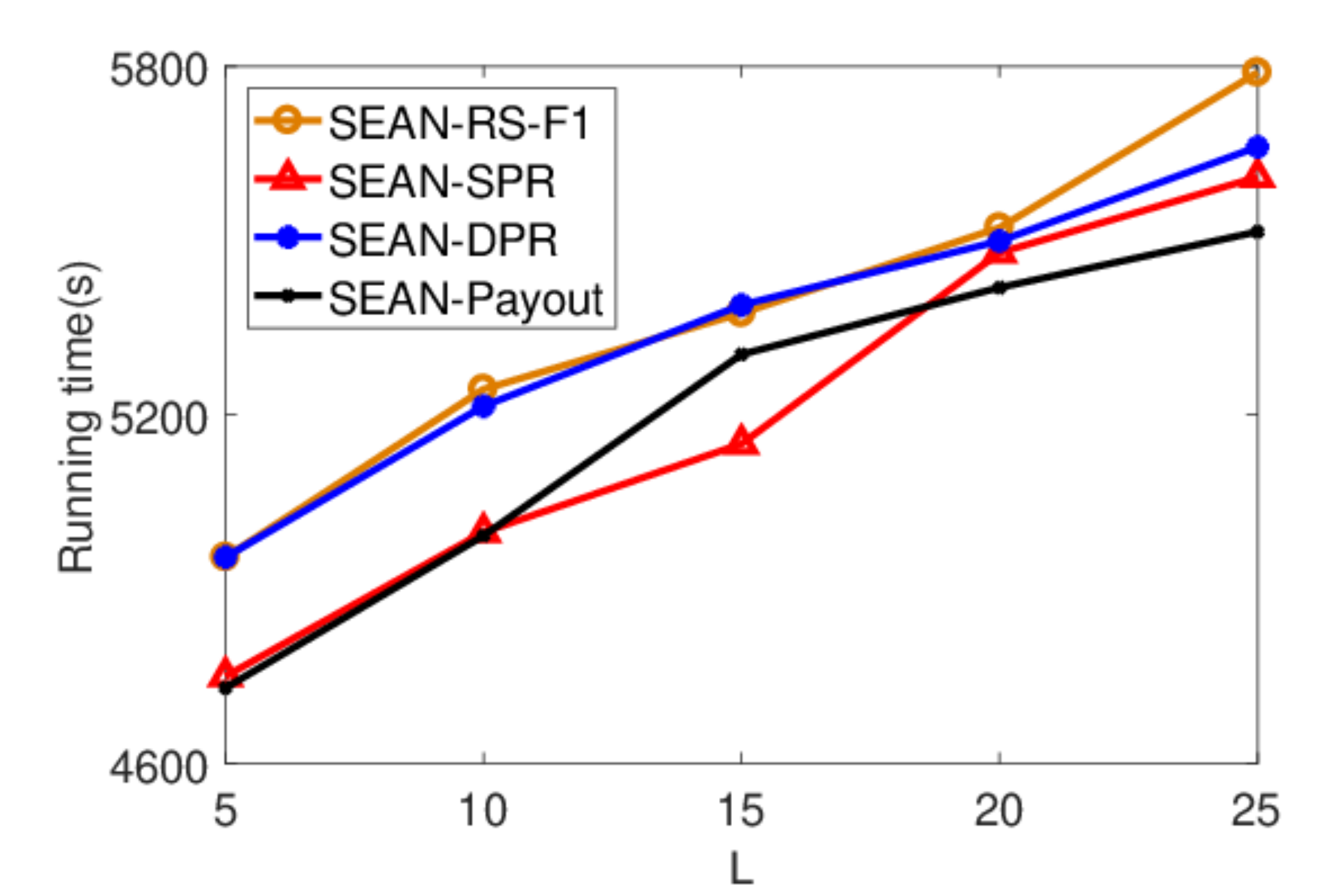}}
    \subfigure[Beam Width $B$.]{\includegraphics[width=0.23\textwidth]{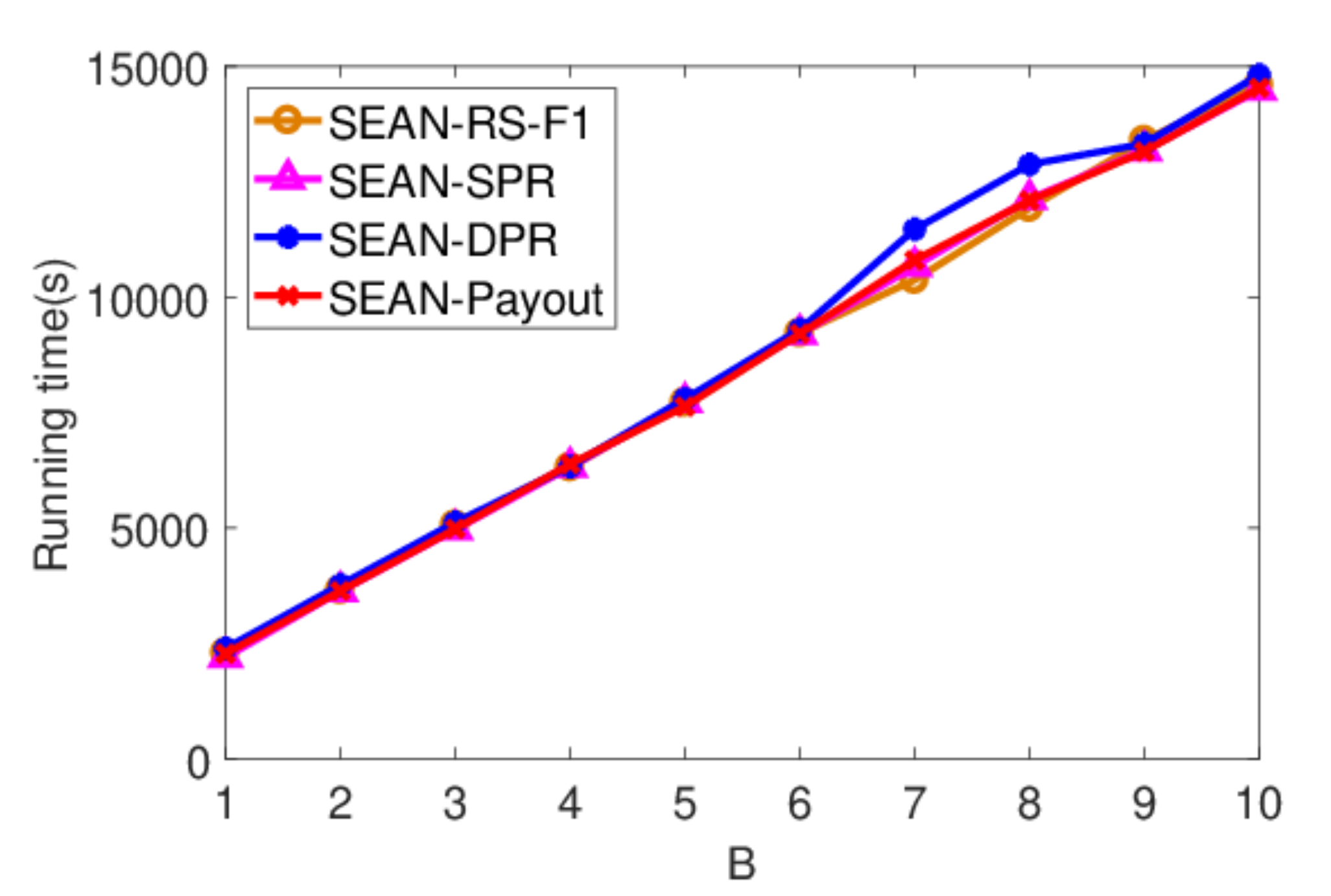}}
    \vspace{-0.1in}
    \caption{Running time w.r.t. $L$ and $B$.}
    \label{fig-scalability}
    \vspace{-0.2in}
\end{figure}

\subsection{Scalability}
The training complexity of SEAN is dominated by the search depth $L$ and the beam width $B$ as explained in Section \ref{sec-complexity}. 
As shown in Figure~\ref{fig-scalability}, the time consumption of SEAN with four different exploitation methods is almost linear to $L$ and $ B$. 
    
\section{Related Work}
\label{sec-related}
For content recommendation, both content based approaches and CF based approaches have been studied.
Content based approaches use a user's historical reading contents to represent a user, so they are naturally personalized models. 
Typical models that recommend documents include~\cite{son2013location,wang2017dynamic,wang2018dkn}.
CF based approaches, including traditional ones~\cite{das2007google,hu2008collaborative,koren2008factorization,koren2009matrix,liu2010personalized,rendle2012factorization,lee2016llorma,zhao2017sloma} and deep learning based models~\cite{wang2015collaborative,he2017neural,li2017collaborative,sun2018attentive,chen2019social}, are also usually called personalized recommendation, as they consider a user's personal preference based on the user's behaviors or actions on the platform. 
There have been many approaches also incorporating social information into CF models, which are usually called social regularization~\cite{jamali2010matrix,ma2011recommender,ye2012exploring,yang2012circle,zhao2017sloma,sun2018attentive}.
More recently, \cite{chen2019social,Song0WCZT19} propose to use social attention instead of regularization to further improve the way of using social information in CF models.
Our work goes beyond social regularization of a personalized model. We propose to use a social exploration mechanism to attend to higher-order friends. In this way, we can find a good balance of using trusted users and explore more users rather than one-hop friends.


Exploitation-Exploration of items is also a hot topic in RS field~\cite{JoachimsFM97,RadlinskiKJ08,li2010contextual,wang2014exploration,liebman2017designing,zheng2018drn,mcinerney2018explore,chen2018large}.
Exploring more items can introduce more diversity in recommendation results.
However, they still only use the click-through rate (CTR) to evaluate their models.
That means most of them still focus on optimizing the performance of recommendation, which only benefits consumers and the platform.
Moreover, they are still working on traditional user-item based collaborative filtering settings. 
There is a lack of studies on content recommendation and focusing on the creators of the contents.
To our knowledge, we are the first work that considers using the Gini coefficient combined with F1 as a core metric to evaluate different recommendation algorithms.
Some existing work, such as~\cite{SalganikDodds}, has used the Gini coefficient to evaluate Matthew's Effect of a RS. However, they have not simultaneously considered recommendation performance.

In addition to recommendation algorithms, mechanism design, e.g.,~\cite{WeiMJ05}, is considered as an orthogonal perspective to improve an RS. The related studies to improve the item equality have been shown in~\cite{AbeliukBHHL17,BerbegliaH17}.

\section{Conclusions}
\label{sec-conclusion}

In this paper, we present a model that goes beyond personalization by exploring higher-order friends in a social network to help content recommendation.
In the model design and the exploration design, we consider the effects for both content creators and consumers in the social media platform.
This can benefit the platform to attract more innovative content creators and encourage more interactions between the creators and consumers.
We use datasets derived from a decentralized content distribution platform, Steemit, to evaluate our proposed framework.
Experimental results show that we can improve both creator's equality and consumer's satisfaction of recommendation results.

\section*{Acknowlegement}
This paper was done when the first author was an intern at WeBank AI Department.
The authors of the paper were also partially supported by the Early Career Scheme (ECS, No. 26206717) from Research Grants Council in Hong Kong and WeChat-HKUST WHAT Lab. We also thank Intel Corporation for supporting our deep learning related research. 

%
\bibliographystyle{plainnat}
\bibliography{sample-bibliography}

\clearpage
\appendix

\section{MCTS based Beam Search Algorithm}
\label{sec:appendexi-bmcts}
We show the MCTS based beam search algorithm in Algorithm~\ref{alg-bmcts} and how the selected friends are used in the SEAN model training in Algorithm~\ref{alg-sean}.
We initialize paths for users by randomly selecting users in the social graph and set users' explored times by the times they selected as friends. For each user in the $t$-th day training, we first explore $B$ sets of friends by MCTS strategy and update the explored times for these selected friends, consequently updating the exploration values $U_t(v)$ of them. These $B$ sets of friends are incorporated with the user as input to the RS model. Then we update $Q_t(u)$ of the target user. The updated results are used for the $t+1$-th day training.

    

\begin{algorithm}[htb]
    \caption{SelectFriends($u, B, L$).}
    \label{alg-bmcts}
    \begin{algorithmic}[1]
    \REQUIRE{target user $u$, beam width $B$, path length $L$}
         \ENSURE{$\big\{\mathcal{F}_b(u)\big\}_{b=1}^B$.}
     \STATE{\textbf{Initialization:} \\
     $\big\{\mathcal{F}_b(u)\big\}_{b=1}^B$: $\mathcal{F}_b(u)$ records the $b$-th path starting from $u$;\\
     $UCB1(v)$: UCB1 score for user $v$ according to Eq.~\eqref{eq-ucb1};\\
     $\big\{T_b(u)\big\}_{b=1}^B$: $T_b(u)$ records the sum of UCB1 scores of the $b$-th path for user $u$ during beam search;\\ 
    $\Delta_b$:the neighbours of the tail node of the path $\mathcal{F}_b(u)$ during beam search;
    }
    
    \WHILE{$k = 0,1,2,\cdots$, L:}
		\STATE{ $\mathcal{H}= \bigcup_{b=1}^B\big\{UCB1(v) + T_b(u), v\in \Delta_b\big\}$;}
		\WHILE{$b = 1,2,\cdots$, B:}
			\STATE{$v = \argmax_v\mathcal{H}$ ;}
			\STATE{$\mathcal{F}_b(u) \leftarrow \mathcal{F}_b(u) \bigcup v;$}
			\STATE{$\mathcal{H}\leftarrow \mathcal{H}\setminus (UCB1(v) + T_b(u))$;}
		\ENDWHILE
    \ENDWHILE
    \end{algorithmic}
\end{algorithm}


\begin{algorithm}[htb]
    \caption{SEAN.}
    \label{alg-sean}
    \begin{algorithmic}[1]
        \FOR{$t = 1,2,\cdots$}
         \FOR{$u \in \mathcal{U}$}
             \STATE{${\{\mathcal{F}_i(u)}\}^B_{b=1}$ = SelectFriends(u, B, L)}
             \FOR{$b=1,2,\cdots,B$}
             \STATE{Train SEAN with $\mathcal{F}_b(u)$ for $u$;}
             \FOR{$v \in \mathcal{F}_b(u)$}
                 \STATE{$N_t(v) \leftarrow N_t(v) + \frac{1}{B}$;}
             \STATE{Update $U_t(v)$ according to Eq.~\eqref{eq-ucb-u};}
             \ENDFOR
                \STATE{Update $Q_t(u)$ for user $u$ according to Section ~\ref{sec-mcts-exploitation};}
            \ENDFOR
            \ENDFOR
        \ENDFOR
    \end{algorithmic}
\end{algorithm}

\section{Experimental Settings}
\label{sec:appendix-settings}
In this section, we present the detailed configuration of the SEAN model in different experiments.
\subsection{More Experimental Settings}
\label{app-sec-setup}
In this part, we present the hyper-parameters we used for training SEAN when comparing with other baselines. The configuration is shown in Table \ref{sec-tb-hp}. 

\begin{table}[H]
\caption{Hyper-parameters of SEAN when comparing with other baselines.}
\label{sec-tb-hp}
{\footnotesize
\begin{tabular}{c|c}
\toprule
\textbf{Hyper-paramter} & \textbf{Value} \\
\midrule
train:validation & 9:1 \\
beam width $B$ & 3 \\
search depth $L$ & 10 \\
trade-off constant $\lambda$ & 1 \\
max number of sentence & 30 \\
max number of word & 100 \\
word embedding size & 300 \\
PageRank $\alpha$ for SPR \& DPR & 0.9 \\
threshold for F1 & 0.5 \\
user embeeding size \& hidden size $h$ & 64 \\
number of CNN kernels $K$ & 6\\
convolutional window size $c$ for $K$ kernels & 1,2,3,4,5,6\\
epochs for training per day & 3 \\
batch size & 128 \\
dropout rate & 0.2 \\
\bottomrule
\end{tabular}
}
\end{table}

\begin{figure*}[t]
    \subfigure[Hidden Size $h$.]{\includegraphics[width=1\textwidth]{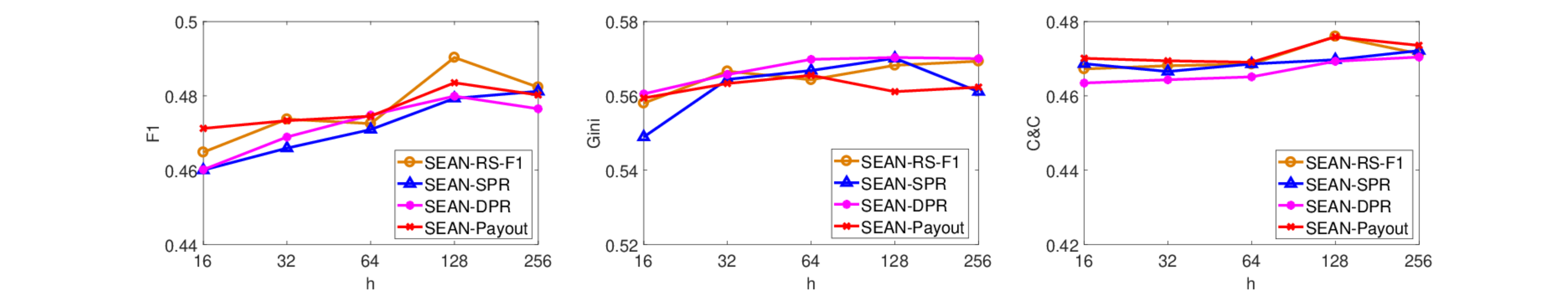}\label{fig-ps-d}}\\
    \subfigure[Number of kernels $K$ in CNN.]{\includegraphics[width=1\textwidth]{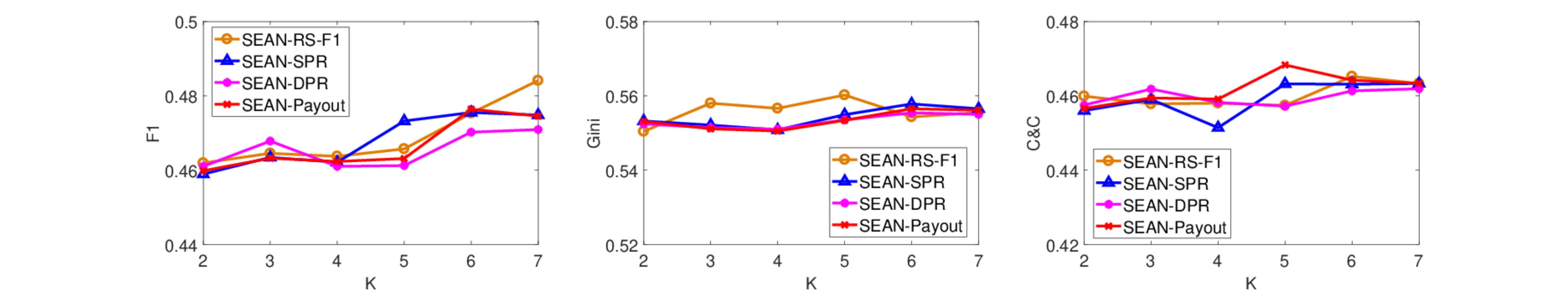}\label{fig-ps-e}} \\
\vspace{-0.1in}
    \subfigure[Filter size $r$.]{\includegraphics[width=1\textwidth]{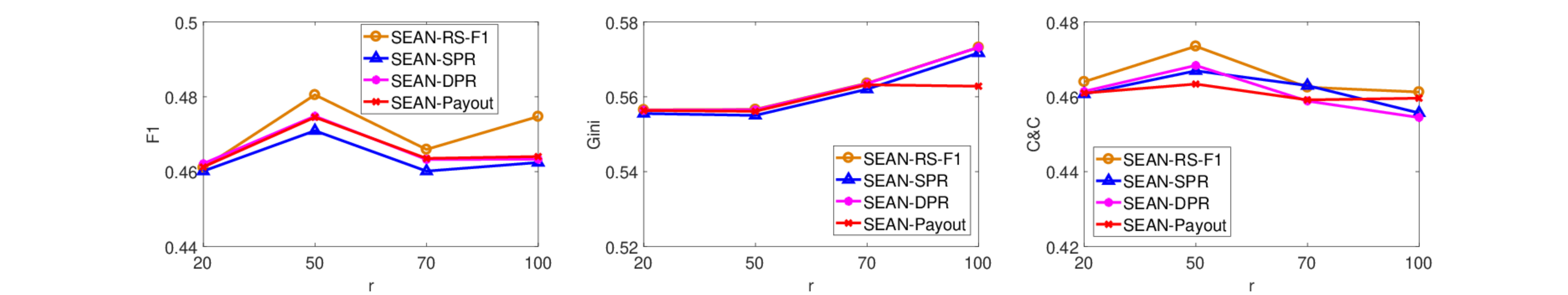}\label{fig-ps-f}}
\vspace{-0.1in}
    \caption{Hyper-parameter sensitivity analysis using the samples from Steemit-English during the first 100 days.}
\vspace{-0.1in}
    \label{app-fig-param}
\end{figure*}

\subsection{Initialization Details}
\label{sec:appendix-initialization}
\textbf{Kernel Initialization.}
Initializers for the kernel weights matrices in CNN \& RNN are Xavier normal initializer:$\text{Var}(W) = \frac{2}{n_\text{in} + n_\text{out}}$,
where $W$ is the initialization distribution for the neuron in CNN \& RNN, $n_{in}$ is the number of neurons feeding into it, and $n_{out}$ is the number of neurons the result is fed to.
Besides, the user embedding matrices $\bA$ and $\bA'$ are initialized by random normal that generates variables in matrices with a normal distribution.

\textbf{Word Embedding.}
Pre-trained word embeddings for Steemit-English and Steemit-Spanish are formulated by performing GloVe \citep{pennington2014glove} on Wikipedia\footnote{https://dumps.wikimedia.org/} and Spanish Billion Word Corpus\footnote{http://crscardellino.github.io/SBWCE/}, respectively. The dimension of the pre-trained embedding is 300. The index of OOV (Out-of-vocabulary) word is set to 0.

\textbf{Path Initialization.}
Before MCTS based strategy updating each user's paths, we need to provide the initialized paths to them. Here we test two random based methods for path initialization:
randomly selecting a set of users on the social platform as users' friends (Random Select) and applying random walk based strategy to generate $B$ paths. From Table~\ref{tab:path-initialization}, we find randomly selecting $L$ users as the target user's friends can achieve slightly better results. 
\begin{table}[ht!]
\caption{Comparison of path initialization strategies.}
\vspace{-0.1in}
{\footnotesize
\begin{tabular}{c|c|c|c}
\toprule
 & F1 & Gini & C\&C \\
 \midrule
 Random Select & \textbf{0.4769$\pm$0.46} & \textbf{0.6178$\pm$0.24} & \textbf{0.4243$\pm$0.33}\\
Random Walk & 0.4667$\pm$0.22 & 0.6134$\pm$0.27 & 0.4221$\pm$0.24 \\
\bottomrule
\end{tabular}
}
\label{tab:path-initialization}
\end{table}

\subsection{Experimental Environment}
The models are implemented using Keras 2.2.4 with Tensorflow 1.31.1 as the backend, based on CUDA 10.0 using single GPU, TITAN Xp, and are tested on Linux (CentOS release 6.9), Python 3.7 from Anaconda 4.6.14.

\section{More Hyper-parameter Sensitivity}
\label{sec:appendix-parameter}
SEAN involves a number of hyper-parameters. In this subsection, we evaluate how different choices of other hyper-parameters affect the performance of SEAN described in Section
\ref{sec-san}. Expect for the parameter being tested, all other parameters are set as introduced in Section \ref{sec-setup}.

\textbf{Hidden Size and User Embedding Size $h$.}
We first investigate how the hidden size $h$ affect the performance by testing $h$ in set $\left \{ 20,50,70,100 \right \}$. The results are shown in Figure \ref{fig-ps-d}, from which we can observe that all four models obtain best F1 when $h=128$. Changing $h$ does not affect too much Gini scores. The trend of C\&C is similar to the trend of F1, also getting the best result when $h=128$.

\textbf{The number of kernels $K$ and sizes of filters $r$.}
We investigate the number of kernels $K$ and the choice of filter sizes $r$ for CNN in SEAN. As shown in Figure \ref{fig-ps-e}, the F1 score generally increases as the number of kernels $K$ with different convolutional windows $g$ gets larger, since more kernels are able to capture long-distance patterns in sentences. Due to the limitation of time and memory, we do not further enlarge the $K$.  Meanwhile, the influence of $K$ on Gini is smaller than on F1. SEAN-F1-RS performs best on F1 while performs worst on Gini.  We can get the best C\&C for all the proposed models except SEAN-RS-F1 when $K=6$.
Likewise, we can observe similar rules for the filter size $r$, shown in Figure \ref{fig-ps-f}: a small filter size cannot capture more local patterns in sentences, while a too large filter size may easily suffer from overfitting. The Gini increases with the increasing of filter size $r$. The best C\&C results are obtained when $r=50$.

\section{Prediction F1 Over Time}
\label{sec:appendix-overtime}
\begin{figure}[h]
\centering
\includegraphics[width=0.4\textwidth]{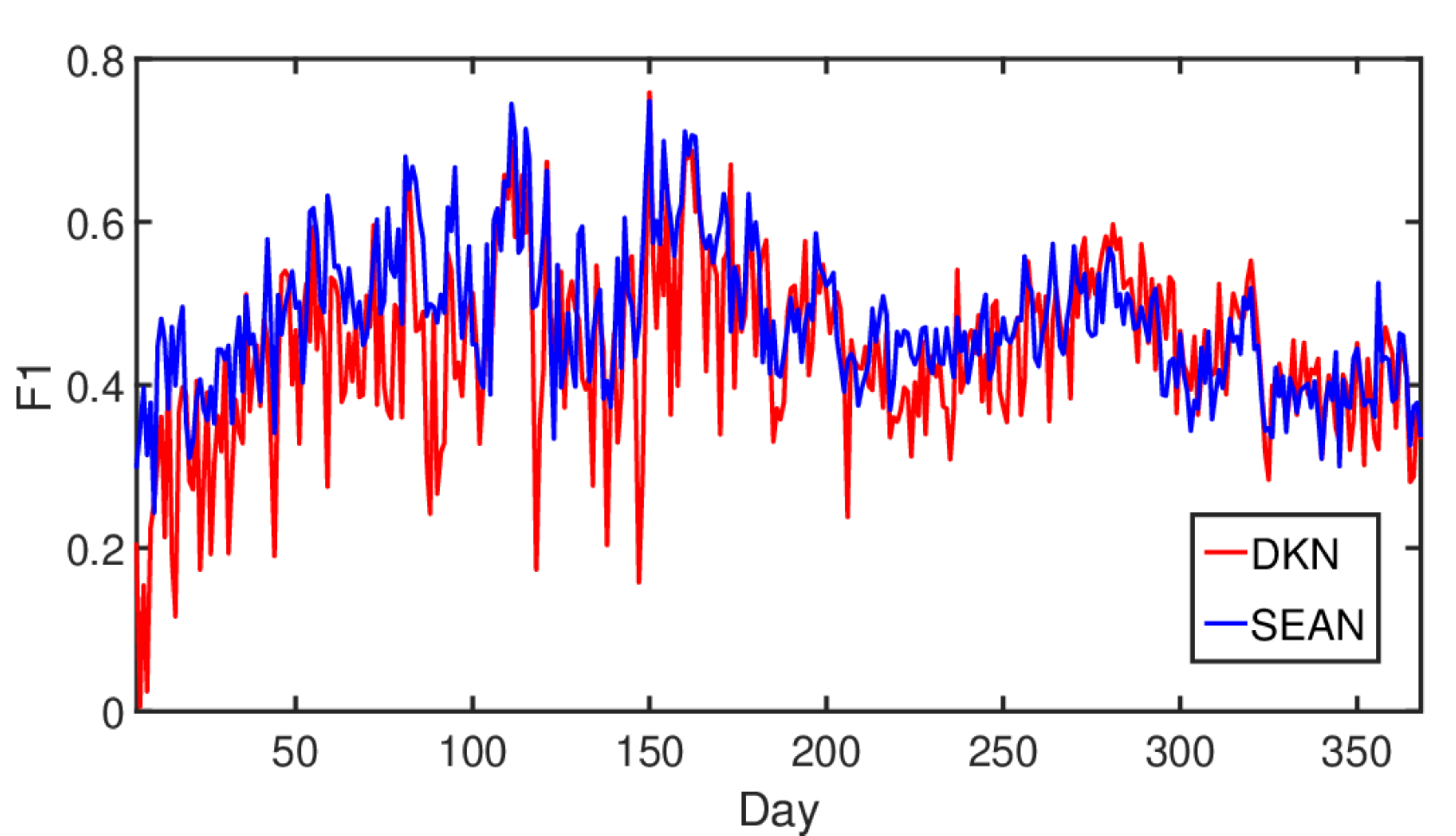}
\vspace{-0.1in}
\caption{Comparison for 368 days.}
\vspace{-0.1in}
\label{baseline-comparison}
\end{figure}

We show the prediction results over 368 days in Figure~\ref{baseline-comparison}.

\end{document}